\documentclass[11pt]{article}
\usepackage[english]{babel}
\usepackage[utf8]{inputenc}
\usepackage{fancyhdr}
\usepackage{graphicx}
\usepackage{float}
\usepackage[paperheight=29.7cm,paperwidth=20.99cm,left=3.0cm,right=3.0cm,top=2.5cm,bottom=2.5cm]{geometry}

\title{The Grass of the Universe: Rethinking Technosphere, Planetary History, and Sustainability with Fermi Paradox}

\author{Luk\'a\v{s} Likav\v{c}an$^{a}$$^{,}$$^{b}$$^{*}$ \\\\
        \small $^{a}$Research Affiliate, Center for AI and Culture, NYU Shanghai \\
        \small $^{b}$Researcher, Antikythera, Berggruen Institute \\\\
        \small $^{*}$\tt{Email: ll5137@nyu.edu, ORCID: 0000-0002-6957-9079}\\\\
}
\date{11 January 2025}

\begin{document}

\maketitle

\pagestyle{fancy}
\fancyhead[L]{\textit{Submitted to Cosmos \& History}}
\fancyhead[R]{\textit{Luk\'a\v{s} Likav\v{c}an}}
\fancyfoot[L]{\textbf{Article preprint version}}

\noindent{\textbf{Abstract: }SETI is not a usual point of departure for environmental humanities. However, this paper argues that theories originating in this field have direct implications for how we think about viable inhabitation of the Earth. To demonstrate SETI’s impact on environmental humanities, this paper introduces the Fermi paradox as a speculative tool to probe possible trajectories of planetary history, especially the “Sustainability Solution” proposed by Jacob Haqq-Misra and Seth Baum. This solution suggests that sustainable coupling between extraterrestrial intelligence and their planetary environments is the major factor in the possibility of their successful detection by remote observation. By positing that exponential growth is not a sustainable development pattern, this solution rules out space-faring civilizations colonizing solar systems or galaxies. This paper elaborates on Haqq-Misra's and Baum's arguments and discusses speculative implications of the Sustainability Solution, thus rethinking three concepts in environmental humanities: technosphere, planetary history, and sustainability. The paper advocates that (1) the technosphere is a transitory layer that shall fold back into the biosphere; (2) planetary history must be understood in a generic perspective that abstracts from terrestrial particularities; and (3) sustainability is not a sufficient vector of viable human inhabitation of the Earth, suggesting instead habitability and genesity as better candidates.\\


\section{Introduction}

Search for Extraterrestrial Intelligence (SETI) refers to a range of efforts in contemporary astronomy, focused on finding traces of intelligent life elsewhere in the universe.\footnote{ Steven J. Dick, Space, Time, and Aliens: Collected Works on Cosmos and Culture (Cham: Springer, 2020), 33–34.} Over seven decades of its existence as a more or less formalized field of inquiry, its key concepts – such as the Fermi paradox, Kardashev scale, or Drake equation – achieved great popularity outside of the narrowly scientific discourses, becoming frequent tropes in sci-fi literature or popularization of space exploration. Furthermore, one can view SETI as an important philosophical inquiry – the highly speculative nature of theorizing about extraterrestrial forms of intelligent life combines methods and findings from astrobiology or exoplanet research with sociological, economic, or anthropological assumptions, ultimately asking fundamental questions about the nature of life or human existence on the cosmic scale.\footnote{ Sara Imari Walker, Life as No One Knows It: The Physics of Life’s Emergence (New York: Riverhead Books, 2024), 225–28.} For example, Milan Ćirković has argued that the Fermi paradox – that the vastness of the cosmos renders the absence of any detectable traces of intelligent life elsewhere in the universe highly counter-intuitive – is a ``successful provocation" that places the human species into ``the context of \textit{independently evolved minds}"\footnote{ Milan M. Ćirković, The Great Silence: Science and Philosophy of Fermi’s Paradox, First edition (Oxford: Oxford University Press, 2018), 282; 266.} (i.e. other intelligent life-forms, sometimes referred to as \textit{sophonts}).\footnote{ I will use the term sophonts interchangeably alongside expressions such as ``extraterrestrials", ``extraterrestrial intelligence" (and its abbreviation ``ETI"), or ``planetary community", to avoid colonial connotations of the word ``civilization". See Sonya Atalay et al., ``Indigenous Studies Working Group Statement," American Indian Culture and Research Journal 45, no. 1 (January 1, 2021): 11–12, https://doi.org/10.17953/AICRJ.45.1.ATALAY\_ETAL.} Fermi paradox thus represents a final step in the Copernican process of removing humans from the center of the world, and its successful resolution may lead to acknowledging a rather unremarkable nature of humanity in the galactic (or even larger) context.\footnote{ Ćirković, The Great Silence, 266.} Beyond that, SETI poses many important social, cultural, or ethical questions: What is the moral status of extraterrestrial life? Is communication with aliens possible at all, and if yes, is attempting the first contact a good idea? What would be the consequences of a successful detection of extraterrestrial intelligence on the world’s religions or international politics?

This article highlights the philosophical import of SETI – and particularly of the Fermi paradox as the central tenet of the field – from the vantage point of environmental humanities, since it motivates substantial reconceptualization of technology, history, and sustainability on a planetary scale. In particular, the article traces the philosophical consequences of the so-called \textit{Sustainability Solution to the Fermi Paradox}, as originally proposed by Jacob Haqq-Misra and Seth Baum.\footnote{ Jacob D. Haqq-Misra and Seth D. Baum, ``The Sustainability Solution to the Fermi Paradox" (arXiv, June 2, 2009), http://arxiv.org/abs/0906.0568.} This solution implies that the apparent lack of any observable extraterrestrial activity may be caused by the ecological unsustainability of exponential economic growth, which would eventually enable the extraterrestrials to colonize the galaxy and thus leave plenty of evidence of their existence.\footnote{ Haqq-Misra and Baum, 47.} Beyond rendering the existence of galaxy-colonizing ETI unlikely, the Sustainability Solution supports the dissolution of the boundary between the natural and the artificial, thus suggesting to following integrative philosophical conceptualizations of the planetary environment and its evolution, such as the notion of \textit{technosphere}.\footnote{ P. K. Haff, ``Technology as a Geological Phenomenon: Implications for Human Well-Being," Geological Society, London, Special Publications 395, no. 1 (2014): 301–9, https://doi.org/10.1144/SP395.4.} It also projects constrained topology of viable futures for any intelligent species (humans included) when it comes to their sustainable coupling with the planetary environment and suggests a reconceptualization of sustainability along the lines of non-human-centric normative principles, such as \textit{habitability} or \textit{genesity}.\footnote{ Wallace Arthur, The Biological Universe: Life in the Milky Way and Beyond, 1st ed. (Cambridge University Press, 2020), https://doi.org/10.1017/9781108873154; Michael L. Wong et al., ``Searching for Life, Mindful of Lyfe’s Possibilities," Life 12, no. 6 (May 25, 2022): 783, https://doi.org/10.3390/life12060783.} These implications of the Sustainability Solution are discussed in the article to reinforce ongoing discussions among the disciplines that fall under the umbrella term of ``environmental humanities", especially surrounding the recent surge of \textit{planetary thinking}.\footnote{ Gayatri Chakravorty Spivak, ``Imperative to Re-Imagine the Planet," in An Aesthetic Education in the Era of Globalization, ed. Gayatri Chakravorty Spivak (Cambridge, MA: Harvard University Press, 2012), 335–50; Dipesh Chakrabarty, ``The Planet: An Emergent Humanist Category," Critical Inquiry 46, no. 1 (September 2019): 1–31, https://doi.org/10.1086/705298.} This discussion unfolds throughout the paper, following a straightforward structure. Chapter 2 introduces the readers to the Fermi paradox, whereas Chapter 3 examines the Sustainability solution proposed by Haqq-Misra and Baum. The next chapters then expose and discuss the three major philosophical implications: Chapter 4 addresses the reconceptualization of technology through the notion of the \textit{technosphere}, Chapter 5 does a similar job concerning the notion of \textit{planetary history}, and Chapter 6 reconceptualizes sustainability from the vantage point of astrobiological notions of habitability and genesity. Finally, Chapter 7 concludes with a summary of the argument, proposing some avenues for further research.

\section{Fermi paradox}

According to Steven J. Dick, the origins of modern SETI can be traced back to the article on interstellar communication by Cocconi and Morrison (1959).\footnote{ Dick, Space, Time, and Aliens, 48.} The authors argue that electromagnetic waves are currently the most likely means of interstellar communication and speculate about the optimal broadcasting frequencies that would be used by a community of intelligent beings to send a message to our solar system.\footnote{ Giuseppe Cocconi and Philip Morrison, ``Searching for Interstellar Communications," Nature 184, no. 4690 (September 1959): 844–45, https://doi.org/10.1038/184844a0.} One year later, Frank Drake initiated Project Ozma at the National Radio Astronomy Observatory (NRAO) in Green Bank – the first systematic effort to intercept potential alien radio waves – by using a radiotelescope to target two Sun-like stars: Tau Ceti and Epsilon Eridiani. Following these attempts, Drake convened a conference at Green Bank in the autumn of 1961, which finally inaugurated SETI as an astronomical research field.\footnote{ Dick, Space, Time, and Aliens, 82–83.} As the story goes, in preparation for this conference, Drake jotted down an equation that would estimate the number of radio-communicating extraterrestrial communities in our galaxy, originally meant to organize the meeting’s agenda:

\vspace{1\baselineskip}
\begin{equation}
N = R_{\ast }\cdot f_{p}\cdot n_{e}\cdot f_{l}\cdot f_{i}\cdot f_{c}\cdot L 
\end{equation}

\vspace{1\baselineskip}
\noindent The total number of such extraterrestrial communities (\textit{N}) is put in this equation in relationship with the star-formation rate in the Milky Way (\( R_{\ast }\)), the fraction of stars with orbiting exoplanets (\textit{f\textsubscript{p}}), the average number of habitable planets per solar system (\textit{n\textsubscript{e}}), the fraction of habitable planets with life (\textit{f\textsubscript{l}}), the fraction of inhabited planets with intelligent life (\textit{f\textsubscript{i}}), the fraction of extraterrestrial communities with sufficient technological capacities for interstellar communication (\textit{f\textsubscript{c}}), and the average lifetime of these technologically advanced communities (\textit{L}).\footnote{ Ćirković, The Great Silence, 225–26.}

\ \ The standard assumption behind this equation is that \textit{N }must be greater than zero and that it may be even a very large number. As such, Drake equation then formalizes the intuition of Enrico Fermi, who at one of the lunches with his colleagues at Los Alamos in 1950 pondered upon the seemingly unexplainable lack of any signs of extraterrestrial activity on Earth, given the apparent vastness of our galaxy, both in terms of its space and in terms of its deep history. William Newman and Carl Sagan later rephrased the paradox in the form of a question: 

\vspace{1\baselineskip}
\noindent``In a galaxy with $\sim$10\textsuperscript{11} stars; with planets apparently abundant and the origin of life seemingly requiring very general cosmic circumstances; with the selective advantage of intelligence and technology obvious and with billions of years available for evolution, should not extraterrestrial intelligence be readily detectable?"\footnote{ William I. Newman and Carl Sagan, ``Galactic Civilizations: Population Dynamics and Interstellar Diffusion," Icarus 46, no. 3 (June 1981): 293, https://doi.org/10.1016/0019-1035(81)90135-4.}

\vspace{1\baselineskip}
\noindent According to their estimates, an activity of extraterrestrial communities with lifespans at least around 20 million years should be observable in the immediate vicinity of our solar system.\footnote{ Newman and Sagan, 318–19.} While viewed from the vantage point of human history, this still amounts to an extremely large portion of time, in the context of the history of our galaxy – which is more than 13 billion years old – 20 million years is just a brief episode. Hence, given the time available for extraterrestrial species to invest in galactic expansion, it is surprising there are no detectable signs of such expansion around us.

\ \ Milan Ćirković offers a more detailed philosophical analysis of the Fermi paradox in his book \textit{The Great Silence}. He distinguishes between several versions of this paradox: \textbf{ProtoFP} (which corresponds to Fermi’s original question about the lack of extraterrestrial activity on Earth), \textbf{WeakFP} (which expands the lack of extraterrestrial activity from Earth to the whole Solar System), and \textbf{StrongFP} (that ultimately concerns the lack of detectable traces of extraterrestrial intelligence to the whole observable universe), plus one special formulation – \textbf{KardashevFP}: ``There is no Kardashev’s Type 3 civilization in the Milky Way."\footnote{ Ćirković, The Great Silence, 4–12.} For this paper, I will use the formulation of the Fermi paradox that Ćirković refers to as \textbf{StrongFP}:

\vspace{1\baselineskip}
\noindent``The lack of any intentional activities or manifestations or traces of extraterrestrial civilizations in our past light cone is incompatible with the multiplicity of extraterrestrial civilizations and our conventional assumptions about their capacities."\footnote{ Ćirković, 10. ``Light cone" is a term that originates in the special theory of relativity. In this case, it represents a spatiotemporal region of all the past events in the universe currently observable by humans on Earth.}

\vspace{1\baselineskip}
\noindent As noted by Ćirković, any good solution to the Fermi paradox relies on attacking its hidden assumptions, starting from those about the origins of life, the frequency of its occurrence in the universe, the number of habitable exoplanets or even the history and nature of our galaxy, and ending with sociological, economic, or cultural assumptions about the behavior of the sophonts.\footnote{ Ćirković, 265.} Once we begin to closely scrutinize these assumptions, they may reveal biases or constraints intrinsic to their construction. For example, the assumption that the evolution of life should be relatively common in the universe may turn out to be unsupported, because humans can currently study only one biosphere from which to infer general conclusions about the nature of life – the terrestrial biosphere. Similarly, even if life in the universe were relatively common, there is no guarantee that the sequence of evolutionary steps is easily reproducible, and hence the emergence of intelligent life may still happen rather seldom.

\section{Sustainability Solution to Fermi Paradox}

Considered from the perspective of the potential import for environmental humanities, the most promising category of solutions to the Fermi paradox is the one that tackles economic, sociological, or cultural assumptions about the dynamics of extraterrestrial communities of intelligent beings. Sustainability Solution to Fermi paradox – as presented by Jacob Haqq-Misra and Seth Baum – represents a good example of this category since it explicitly tackles conventional presuppositions about the behavior of the extraterrestrials, manifested in the expectation that their communities will tend to expand outside of their planet/solar system employing advanced technologies, to colonize more and more star systems in the galaxy. As the authors state, the formulation of the Fermi paradox contains a biased presupposition based on the observation of only one planetary community of intelligent species (i.e. humans), which is in turn based on a warped understanding of human history, which assumes that the history unfolds in a progressive series of civilizational, colonial expansions.\footnote{ Haqq-Misra and Baum, ``The Sustainability Solution to the Fermi Paradox," 47.} However, this understanding does not represent human history, neither in terms of its \textit{long durée} nor in terms of the sum total of societies composed of the members of \textit{homo sapiens}.\footnote{ Haqq-Misra and Baum, 47.} Therefore, the authors continue with the corollary that ``[t]he absence of ETI observation can be explained by the possibility that exponential growth is not a sustainable development pattern for intelligent civilizations,"\footnote{ Haqq-Misra and Baum, 49.} and they eventually claim that if ETI exists, it is not detectable in the sense that Fermi paradox expects, i.e. by leaving traces of its expansionist interstellar activity. Instead, the authors hypothesize that ``\textit{exponentially expansive ETI civilization }does not exist,"\footnote{ Haqq-Misra and Baum, 49.} which however does not rule out the existence of non-expansionist ETI. The open question then becomes: How would the economic, sociological, or cultural dynamics of such ETI look like?

To address this question, it is important to first understand the unsustainability of exponential economic growth itself. That this kind of growth is the dead-end for the global economy became one of the major critical insights of contemporary environmental thinking, especially well-documented and formalized by ecological economics.\footnote{ Herman E. Daly and Joshua C. Farley, Ecological Economics: Principles and Applications, 2nd ed (Washington, DC: Island Press, 2010).} For example, Nicholas Georgescu-Roegen warned that economic growth inevitably leads to resource depletion, which in the long run entails impediments of economic growth, or even outright economic collapse, given that the available resources for the human planetary economy are constrained.\footnote{ Nicholas Georgescu-Roegen, ``Energy and Economic Myths," Southern Economic Journal 41 (1975): 363–67.} He supported his conclusion with extensive integration of thermodynamics into economic thinking – especially the second law of thermodynamics, which states that the waste, unusable energy in any system tends to increase with time.\footnote{ Whereas entropy is an index of this waste energy, see Georgescu-Roegen 351.} More recently, it became customary in ecological economics to view the economy as a \textit{metabolism}\footnote{ Mario Giampietro, Kozo Mayumi, and Alevgul H. Sorman, The Metabolic Pattern of Societies: Where Economists Fall Short, Routledge Studies in Ecological Economics 15 (Abingdon, Oxon: Routledge, 2012), 177, https://doi.org/10.4324/9780203635926.} – a model that implicitly assumes homeostatic regulation (rather than exponential expansion) of material, energetic, and informational throughput for the sake of the maintenance of the metabolic process in place.\footnote{ Margaret A. Boden, ``Is Metabolism Necessary?," The British Journal for the Philosophy of Science 50, no. 2 (1999): 236–37.} Additionally, Will Steffen and his colleagues have provided ample indicators for the rapid exponential growth of the human global economy since the 1950s (so-called \textit{Great Acceleration}), thus documenting the current unsustainable pathway of our planetary community.\footnote{ Will Steffen et al., ``The Trajectory of the Anthropocene: The Great Acceleration," The Anthropocene Review 2 (2015): 81–98.}

While Haqq-Misra and Baum do not refer to this literature, they claim that ``exponentially expansive practices are commonly considered unsustainable,"\footnote{ Haqq-Misra and Baum, ``The Sustainability Solution to the Fermi Paradox," 48.} and they even mention some scholars with economic viewpoints adjacent to those of ecological economics – especially Dietz, Ostrom and Stern, who devise strategies for environmental governance that take into account depletion or regeneration rates of different natural resources.\footnote{ Thomas Dietz, Elinor Ostrom, and Paul C. Stern, ``The Struggle to Govern the Commons," Science 302, no. 5652 (December 12, 2003): 1907–12, https://doi.org/10.1126/science.1091015.} Accordingly, it is reasonable to assume that if the sustainability solution holds, any viable ETI community will tend to maintain its economy in degrowth, post-growth, or steady-state conditions.\footnote{ Lukáš Likavčan and Manuel Scholz-Wäckerle, ``Technology Appropriation in a De-Growing Economy," Journal of Cleaner Production 197 (October 2018): 1666–75, https://doi.org/10.1016/j.jclepro.2016.12.134; Christian Kerschner, ``Economic De-Growth vs. Steady-State Economy," Journal of Cleaner Production 18, no. 6 (April 2010): 544–51, https://doi.org/10.1016/j.jclepro.2009.10.019.} Put in the less technical vocabulary, the sustainability solution to the Fermi paradox then implies that the absence of any detectable extraterrestrial activity is caused by the fact that intelligent life in the universe either self-destroys before reaching the point when its activity can be observable by our instruments or that it is too busy saving its planetary environment to make any spectacular gestures to the potential cosmic audience. The potentiality of frequent self-destruction trajectories of ETI has been recently suggested also by Wong and Bartlett: the \textit{asymptotic burnouts} of ETI communities caused by \textit{singularities}, defined as end states of ``trajectories that are headed towards a state of infinite population and energy usage in a finite amount of time."\footnote{ Michael L. Wong and Stuart Bartlett, ``Asymptotic Burnout and Homeostatic Awakening: A Possible Solution to the Fermi Paradox?," Journal of The Royal Society Interface 19, no. 190 (May 2022): 3–4, https://doi.org/10.1098/rsif.2022.0029.} Despite their hypothesis does not rule out post-burnout ETI communities (however difficult to detect), it still renders expansionist ETI communities highly unlikely. Most importantly, the authors emphasize that even though singularities may be avoided or even recovered from, the repetition of the singularity events, in the long run, increases the likelihood of societal collapse. Despite this conclusion, Wong and Bartlett keep a narrow window open for expansionist ETI, namely those that pass the hypothetical, extremely selective evolutionary point of transition called the \textit{Great Filter}.\footnote{ Robin Hanson, ``The Great Filter - Are We Almost Past It?," September 15, 1998, https://mason.gmu.edu/$\sim$rhanson/greatfilter.html.} 

Returning to Haqq-Misra and Baum’s original paper, their solution to the Fermi paradox has direct implications for ``human civilization management". They introduce the proposition \textbf{Need-SD} – ``human civilization needs to transition to sustainable development to avoid collapse"\footnote{ Haqq-Misra and Baum, ``The Sustainability Solution to the Fermi Paradox," 50.} – adding that it does not have a status of the logically necessary implication of the Sustainability Solution, since it does not say that humans \textit{should} avoid collapse, or that the collapse cannot come by other means than through growth overshoot. Still, according to the authors, \textbf{Need-SD} suggests we must place a ``greater-than-zero chance that the absence of ETI observation is due to the unsustainability of exponential civilization growth patterns."\footnote{ Haqq-Misra and Baum, 50.} For this reason, Sustainability Solution exemplifies how outer space research feeds back into human affairs here on Earth, which Lisa Messeri calls \textit{gestures of cosmic relation}.\footnote{ Lisa Messeri, ``Gestures of Cosmic Relation and the Search for Another Earth," Environmental Humanities 9, no. 2 (November 1, 2017): 327, https://doi.org/10.1215/22011919-4215325.} It well aligns with the maxim set by Freeman Dyson: ``Every search for alien civilization should be planned to give interesting results even when no aliens are discovered."\footnote{ NASA Technosignatures Workshop Participants, ``NASA and the Search for Technosignatures: A Report from the NASA Technosignatures Workshop" (arXiv, 2018), 3, https://doi.org/10.48550/ARXIV.1812.08681.}

\section{Indistinguishable from nature: Rethinking technosphere}

Beyond its potential consequences for Earthlings, the Sustainability Solution to Fermi paradox also challenges current approaches in SETI, especially when it comes to the notion of \textit{technosignatures}, defined as ``\textit{any} sign of technology that we can use to infer the existence of intelligent life elsewhere in the universe" (including those that involve some observable artificial modification of the planetary or cosmic environment).\footnote{ NASA Technosignatures Workshop Participants, 2.} Among these, we can count relatively uncontroversial potential technosignatures such as radio transmission (which points to its artificial origin due to unusual regularities in pattern/frequency) as well as hypothetical grandiose astroengineering projects such as Dyson spheres (megastructures used to harvest stellar energy).\footnote{ NASA Technosignatures Workshop Participants, 4.} According to Haqq-Misra and Baum, the Sustainability Solution constraints the range of probable technosignatures to electromagnetic signals by no-growth or post-collapse ETI communities, graveyard technosignatures, and Solar System SETI (e.g. searching for automated probes sent by ETI hiding in our stellar neighborhood).\footnote{ Haqq-Misra and Baum, ``The Sustainability Solution to the Fermi Paradox," 49–50.} However, with a bit of extra philosophical exegesis, one can take the conclusions of the authors even further, eventually circling back to the human affairs on Earth: Is it possible that hypothetical post-growth ETI communities direct their technological development so that their artificial creations maximally mimic their natural planetary environment, thus ultimately blurring the boundary between nature and culture (frequently attacked across contemporary environmental humanities)?

In this respect, Canadian sci-fi author Karl Schroeder has proposed that ``any sufficiently advanced technology is indistinguishable from \textit{nature}" [italics L.L.] – a variation on Arthur C. Clark’s famous ``any sufficiently advanced technology is indistinguishable from \textit{magic}" [italics L.L.].\footnote{ Karl Schröder, ``The Deepening Paradox," 2011, https://www.kschroeder.com/weblog/the-deepening-paradox.} Extrapolating from the Sustainability Solution, the equivocation of ``sufficiently advanced technology" with ``nature" suggests a trajectory of technical evolution that bends towards convergence with planetary metabolisms by expanding or repeating them, rather than dramatically modifying them.\footnote{ See also Laura Tripaldi, Parallel Minds: Discovering the Intelligence of Materials (Falmouth: Urbanomic, 2022), 11.} Hence, using the words of French philosopher of technology Gilbert Simondon, the mature stage of the evolution of technical objects – what he calls their ``concretization" – is ``analogous to that of natural spontaneously produced objects."\footnote{ Gilbert Simondon, On the Mode of Existence of Technical Objects (Minneapolis, MN: University of Minnesota Press, 2017), 50.} Here, the philosophy of technology also meets with an assumption about the socio-economic organization of ETI. Echoing implications of the proposal by Haqq-Misra and Baum, Michael Wong and Stuart Bartlett also recall Schroeder’s quote, and suggest that ``persistent civilizations that transition through homeostatic awakening may be difficult or impossible to detect"\footnote{ Wong and Bartlett, ``Asymptotic Burnout and Homeostatic Awakening," 8.} – whereas ``homeostatic awakening" is their term for ``fundamental, systemic change that alters its \textit{modus operandi} away from unbounded growth that results in an arbitrarily large demand on energy in a finite amount of time."\footnote{ Wong and Bartlett, 5.} At first glance, such a proposal runs afoul of one of the other cornerstones of SETI – the Kardashev scale, which divides ETI communities according to the amount of energy it can harness from the cosmic environment. Originally proposed by Nikolai Semyonovich Kardashev, the scale distinguishes three technological levels:\\

\vspace{1\baselineskip}
``I – technology level close to the level presently attained on the earth, with energy consumption at $\sim$4 x 10\textsuperscript{19} erg/sec.

II – a civilization capable of harnessing the energy radiated by its own star (for example, the stage of successful construction of a ‘Dyson sphere’); energy consumption at $\sim$4 x 10\textsuperscript{33} erg/sec.

III – a civilization in possession of energy on the scale of its own galaxy, with energy consumption at $\sim$4 x 10\textsuperscript{44} erg/sec."\footnote{ N. S. Kardashev, ``Transmission of Information by Extraterrestrial Civilizations.," Soviet Astronomy 8 (October 1, 1964): 219.}

\vspace{1\baselineskip}
\noindent Whereas Type I may be still difficult to detect, Kardashev assumed that ETI communities belonging to Type II and III would leave unambiguous traces of their activity, in line with standard assumptions of the Fermi paradox (namely the tendency to outward expansion).\footnote{ According to current estimates, Earth’s human planetary community stands roughly at Type 0.7276, with an estimated increase to Type 0.7449 by 2060. Antong Zhang et al., ``Forecasting the Progression of Human Civilization on the Kardashev Scale through 2060 with a Machine Learning Approach," Scientific Reports 13, no. 1 (July 12, 2023): 11305, https://doi.org/10.1038/s41598-023-38351-y.} However, the detectability of Type III can be questioned, at least as far as one imagines such ETI as a megalomaniac spacefaring race of galactic colonialists. Looking back to Carl Sagan’s \textit{Contact }(1985), the argument behind the story implies that a Type III (or even higher-level) sophonts could be capable of astroengineering projects that would be unrecognizable from ``naturally spontaneously produced objects" in the cosmos – i.e. building stars or planets, or even whole universes. Sustainability Solution thus potentially radicalizes Sagan’s speculation, suggesting that the threshold where nature and technology merge lies way ahead of Type III: What if it is already Type I, and potentially even sooner?

One of the offshoots of the Anthropocene debate over the last two decades has been theories preoccupied with the notion of \textit{technosphere}. It denotes a new emerging layer of the Earth system, extending beyond both the geosphere and biosphere, and leading to artificial alteration of these evolutionary older layers (best represented by the disastrous environmental impact of the exponential growth of the global economy in the post-WWII period).\footnote{ Steffen et al., ``The Trajectory of the Anthropocene: The Great Acceleration."} In this concept, thinking about technology as a continuation of nature by other means finds its articulation through the geological agency of the human planetary community. Humans are understood here as members of the biosphere – to the extent to which they are biological organisms – and at the same time ``constructors" of technological objects.\footnote{ The term ``constructor" is used here deliberately to invoke a technical notion of constructors as devised David Deutsch ``Constructor Theory," Synthese 190, no. 18 (December 2013): 4331–59, https://doi.org/10.1007/s11229-013-0279-z.} Moreover, what makes a technosphere a true ``sphere" (not just a collection of technical objects) is its planetary scale and interconnected nature. As Peter Haff puts it:

\vspace{1\baselineskip}
\noindent ``The proliferation of technology across the globe defines the technosphere – the set of large-scale networked technologies that underlie and make possible rapid extraction from the Earth of large quantities of free energy and subsequent power generation, long-distance, nearly instantaneous communication, rapid long-distance energy and mass transport, the existence and operation of modern governmental and other bureaucracies, high-intensity industrial and manufacturing operations including regional, continental and global distribution of food and other goods, and a myriad additional ‘artificial’ or ‘non-natural’ processes [$\ldots$]."\footnote{ Haff, ``Technology as a Geological Phenomenon," 1.} 

\vspace{1\baselineskip}
\ \ From the historical perspective, one of the crucial goals in the successful preservation of a new geological layer is its ability to co-exist with the older ones, which means first and foremost an ability to successfully interface with these layers on both its input side (i.e. the resources it takes from the older layers) and output side (i.e. the waste it disposes back into other layers). Following Haff, a good example here may be the Great Oxidation Event, which was preceded by the evolution of cyanobacteria –microorganisms capable of photosynthesis. While this chapter from the history of our planet presents a case of extremely successful energy extraction based on inputs from atmospheric carbon dioxide and sunlight, it had tragic consequences on its output side: the accumulation of oxygen in the atmosphere led to the mass extinction of anaerobic organisms, for whom oxygen acted as a toxin. The survivors were forced to adapt either by inventing mechanisms to metabolize oxygen or to retreat to remaining patches of anaerobic environments. A similar scenario may be currently underway in the case of the technosphere – its dependence on fossil fuels leads to dumping large quantities of CO\textsubscript{2} into Earth’s atmosphere in an extremely short period, which undermines the conditions of existence of technosphere’s constructors, i.e. humans. Based on this observation, Haff claims that for the technosphere to establish itself as a geological layer proper, it must solve its waste management problem. Given the current improbability of the large-scale carbon drawdown by technological means that would overtake the output rate of carbon emissions, the most likely solution to this problem lies on the input side, i.e. in how the technosphere gains the energy from its environment. As the constructor of the terrestrial technosphere, the human planetary community has a decisive role in this adaptation.

Returning now to the earlier claim that the Sustainability Solution may blur the boundary between nature and culture, a sustainable coupling between technosphere and biosphere may lead to their close alignment in terms of their inputs and outputs, which is consistent with the current lack of observation of any distinct technosignatures proper. Even if the technosphere becomes a regulatory armature of the biosphere, it achieves this function only under the condition that it models the regulated biospheric metabolisms so veritably that it essentially becomes indistinguishable from them.\footnote{ Roger C. Conant and W. Ross Ashby, ``Every Good Regulator of a System Must Be a Model of That System," International Journal of Systems Science 1, no. 2 (October 1970): 89–97, https://doi.org/10.1080/00207727008920220.} Such a statement oscillates between a metaphysical statement about the nature of technology and an epistemological corollary regarding the analytical import of the concept of technosphere. The technosphere seems to be under this light just a transitory geological layer that reintegrates back into its biological ``substrate" down the line, and so the value of this concept lies mainly in distinguishing the temporary vehicle facilitating a ``major transition" in planetary evolution.\footnote{ Hikaru Furukawa and Sara Imari Walker, ``Major Transitions in Planetary Evolution," in The 2018 Conference on Artificial Life (The 2018 Conference on Artificial Life, Tokyo, Japan: MIT Press, 2018), 101–2, https://doi.org/10.1162/isal\_a\_00024.} The technosphere is in this view something that the terrestrial biosphere produces as a sensing, modeling, and regulatory mechanism, not as a suffocating blanket destined to replace or sit on top of organic matter, somehow extinguishing its energetic and creative potential. Hence, the catastrophic scenario of runaway climate change compromising the sustainable existence of a planetary community is not the result of the emergence of technosphere as such, but of its failed reintegration. The Sustainability Solution to the Fermi Paradox\textit{ }thus supports a major evolutionary-historical implication: \textit{successful technospheres fold back into biospheres}. This implication still requires backing by robust theoretical support as well as potential empirical evidence, and for this reason, it is stated in this paper only as a speculative philosophical corollary, not as a hard consequence of the Sustainability Solution.

\section{Topology of any viable future: Rethinking planetary history}

In the most philosophical chapter of his sci-fi novel \textit{Solaris} (1961), Stanislaw Lem lets one of the main characters articulate his vision of the unremarkability of human existence in the cosmic context, summed up by the statement: ``Us, we’re common, we’re the grass of the universe, and we take pride in our commonness, that it’s so widespread, and we thought it could encompass everything."\footnote{ Stanisław Lem, Solaris (Pro Auctore Wojciech Zemek, 2014), 166.} The metaphor of the ``grass of the universe" is central to this paper, as it recognizes the crucial environmental implication of the potential existence of intelligent life elsewhere in the universe – namely that the history of the human planetary community on Earth is not a singular occurrence, but potentially unfolds throughout the cosmos in many permutations, conditioned by the setting of given star system and the inhabited exoplanet(s). The grass as a relatively common category of plants (at least for the geographical regions of Eastern Europe from where both Lem and the author of this article come from) is allegorically linked here with the humans as members of a relatively common category of sophonts (while not being representative of sophonts in general). In turn, the potentially large sample of planetary communities (i.e. the high value of parameters \textit{f\textsubscript{i}} and \textit{f\textsubscript{c}} in the Drake equation) may lead to conclusions regarding the likelihood of certain historical pathways, which implies a constrained space of possible trajectories of planetary history. 

As already discussed in Chapters 3 and 4, Haqq-Misra and Baum conclude that their solution impacts ``human civilizational management", since ``the Sustainability Solution increases the likelihood that human civilization needs to transition towards sustainable development in order to avoid its own collapse."\footnote{ Haqq-Misra and Baum, ``The Sustainability Solution to the Fermi Paradox," 51.} Although Ćirković explicitly disagrees with authors of the Sustainability Solution paper – since he claims that remotely detectable astroengineering projects such as Dyson spheres offer better paths to sustainability than deliberate avoidance of exponential economic growth\footnote{ Ćirković, The Great Silence, 222.} – he reaches the same conclusion when it comes to the consequentiality of SETI research for human future on Earth: what we can know or infer about ETI matters for plotting the space of possible futures we can sustainably inhabit (where ``sustainably inhabit" may simply mean to successfully prolong the value of our planetary community’s \textit{L}-parameter in Drake equation). As he says: ``With a few exceptions, the choice of the optimal solution to Fermi’s problem severely constrains the options for the future evolutionary trajectory of our own, human civilization."\footnote{ Ćirković, 266.}

To better understand the idea of constrained space of historical pathways, one can visualize them akin to the Hertzsprung-Russell diagram (see \textbf{Figure 1}). H-R diagram charts the luminosity of the stars alongside their color spectrum (from white through yellow to red, representing their surface temperature), but it simultaneously contains information about the evolutionary pathways of stars: in the middle of the chart, we find the so-called Main sequence, which represents the most common population of stars in the known universe, including our Sun. Stars on the Main sequence eventually exit this category after they burn all the hydrogen that fuels nuclear fusion, moving upwards the chart towards giant stars with higher luminosity, while also shifting their color spectrum towards red as they become relatively cooler to their Main sequence phase. After a few million years, red giants explode in supernovas and release cosmic material into their neighborhood, while the stars themselves collapse into dwarf stars (the bottom of the diagram), or they turn into non-stellar objects (depending on the star’s mass, it can be a pulsar, neutron star, or black hole).\footnote{ Raymond T. Pierrehumbert, Principles of Planetary Climate, 1st ed. (Cambridge University Press, 2010), 22–24, https://doi.org/10.1017/CBO9780511780783.} Now, in a similar vein, a planet can move on the imaginary chart of its evolutionary history, with each move marking its major evolutionary transition. Although the idea of planetary evolution is mostly used to refer to the planet’s initial formation, one can use the same framework to project further stages of evolution, including the emergence of biospheres and later technospheres.\footnote{ Adam Frank, David Grinspoon, and Sara Walker, ``Intelligence as a Planetary Scale Process," International Journal of Astrobiology 21, no. 2 (April 2022): 47–61, https://doi.org/10.1017/S147355042100029X.} As a result, one can envision a planetary analogy to the H-R diagram as a multi-dimensional plot of possible categories planets may transition to during their lifetime.

\vspace{1\baselineskip}
\begin{figure}[H]
\centering
\includegraphics[width=14.35cm,height=16.33cm]{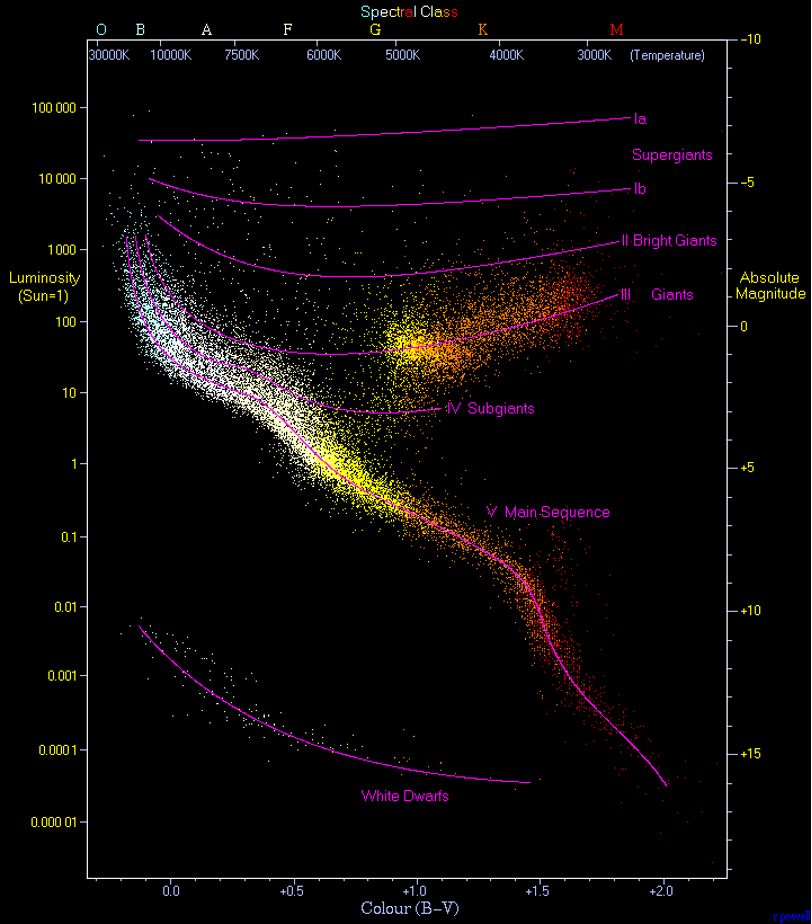}
\caption{Richard Powell, The Hertzsprung Russell Diagram, CC BY-SA 2.5, via Wikimedia Commons\\}
\label{fig:richard_powell_the_hertzsprung_russell}
\end{figure}

The study of planetary history thus becomes possible only in sufficiently broad perspective of astronomy, and morphs into a science of generic, cross-planetary ecology, inaugurating a kind of \textit{comparative planetology of history}.\footnote{ Dipesh Chakrabarty, The Climate of History in a Planetary Age (Chicago; London: The University of Chicago Press, 2021), 67; Luk\'a\v{s} Likav\v{c}an, Introduction to Comparative Planetology (Moscow: Strelka Press, 2019).} In this context, the Fermi paradox and its hypothetical solutions can be seen as probative tools for pruning unviable historical pathways in an experimental medium of astronomical speculation (``unviable" especially for the co-existence between the planet and its community of sophonts). The multi-dimensional plot of possible planetary histories represents a kind of spatial rendering of possible planetary pathways: \textit{topology of planetary history}. This topology diagrams how planets co-evolve with their inhabitants, what major transitions they undertake, or when the historical trajectories lead to a dead end. As an example of such a multi-dimensional plot, consider the diagram presented by Wong and Bartlett depicting space of possible pathways of planetary communities across two major transitions – homeostatic awakening and hypothetical advancement to Kardashev Type III civilization – and featuring narrow bottlenecks transporting the planets towards either of these transitional options.\footnote{ Wong and Bartlett, ``Asymptotic Burnout and Homeostatic Awakening," 7.} The diagram also includes a collapse scenario (asymptotic burnout) and locates the position of the Earth in the overall planetary-historical topology, including its available future pathways.

Concerning the Sustainability Solution, the ecological limits constrain the topology of viable planetary histories to those evolutionary trajectories where the technosphere successfully folds back into the biosphere. In this sense, one can imagine the segment of the topology of planetary history \textit{after} the emergence of the technosphere as a space carved out from the values of different parameters of coupling between technosphere and biosphere (i.e. compatible with the proposition \textbf{Need-SD}), with the objective of continuous inhabitation of the planet by a community of sophonts in mind (i.e. extending Drake’s \textit{L}-parameter). These values may then represent optimal evolutionary pathways repeatable in different permutations across a wider range of planets and their categories diagrammed on our hypothetical plot, which suggests predictable convergences in historical trajectories across similar exemplars of biosphere-technosphere couples.\footnote{ Cyrille Jeancolas et al., ``Is Astrobiology Serious Science?," Nature Astronomy 8, no. 1 (December 28, 2023): 6, https://doi.org/10.1038/s41550-023-02165-9.}

\section{Habitability and genesity: Rethinking sustainability}

One of the remaining open questions to address now is how to better articulate the parameters of the topological diagram within which the viable trajectories of planetary history exist. The Sustainability Solution specifies it quite straightforwardly in the proposition \textbf{Need-SD}: ``[$\ldots$] human civilization needs to transition to sustainable development in order to avoid collapse."\footnote{ Haqq-Misra and Baum, ``The Sustainability Solution to the Fermi Paradox," 50.} But is sustainable development enough? According to the Brundtland Commission, sustainable development can be defined as meeting ``the needs of the present without compromising the ability of future generations to meet their own needs."\footnote{ World Commission on Environment and Development, ``Our Common Future". Report of the World Commission on Environment and Development (New York: UN, 1987).} It establishes some inter-generational balance sheet within the community of \textit{homo sapiens}, but it lacks any clear commitment to boundaries set by the dynamics of the Earth system, despite one may argue these boundaries are implicitly invoked here. Nevertheless, the definition of sustainability unveils its major shortcoming: policy goals based on the principle of sustainability largely target only human well-being, and they see the rest of the planet as mere supporting system.\footnote{ Chakrabarty, ``The Planet," 18.} 

The concept of sustainability can be contrasted to emergent categories that challenge human-centric assumptions behind ecological principles, as Chakrabarty illustrates with the concept of \textit{habitability}. Borrowed from astrobiology, the concept of habitability is very broadly defined as the ability of an environment to sustain Earth-like life.\footnote{ Charles S. Cockell et al., ``Sustained and Comparative Habitability beyond Earth," Nature Astronomy 8, no. 1 (December 28, 2023): 30, https://doi.org/10.1038/s41550-023-02158-8.} This concept is crucial for the search for life in the universe – finding habitable exoplanets is a precondition for a targeted search for signs of life on these candidate worlds. According to Chakrabarty, habitability can replace sustainability as a vector of viable planetary futures, since it guides us to be attentive to ``what makes a planet friendly to the continuous existence of complex life"\footnote{ Chakrabarty, The Climate of History in a Planetary Age, 83.} – i.e. not just human life, as evident from the astrobiological definitions of habitability and habitable zone. Chakrabarty assumes that maintenance habitability is ultimately beneficial to \textit{homo sapiens}, especially when it comes to the critical role of biodiversity for human flourishing.\footnote{ Chakrabarty, 203–4.} Embracing habitability is thus a gesture of ethical generosity – it expands the realm of who matters in terms of preserving Earth’s livability to the whole biosphere. 

Yet, an objection may appear at this point: Given the attempt to collapse the notion of technosphere back into the biosphere in Chapter 4, is the guarantee of habitability for Earth-like life sufficient? This objection is closely tied to an ongoing debate in astrobiology, where the Earth-likeness in the definition of habitability seems to severely constrain the search space for possible forms of alien life. As argued by Sara Walker, the possible chemical configurations of basic building blocks of life – such as nucleotides that constitute RNA and DNA macromolecules – are much larger than the versions we observe here on Earth.\footnote{ Walker, Life as No One Knows It, 169.} At the same time, the ability to transmit information across subsequent generations can be considered a universal property of anything we would call life, whether RNA/DNA-based or not. For this reason, she concurs that when it comes to the chemical makeup of this or that biosphere, the ``general features may be universal, but the specific implementation will be geochemistry dependent."\footnote{ Walker, 168.} In a similar vein, Michael Wong with Stuart Bartlett, Sihe Chen, and Louisa Tierney try to bridge the gap between ``life-as-we-know-it" and ``life-as-we-do-not-know-it" by construing a generic concept of \textit{lyfe} which would capture the main general features of any biosphere’s substrate.\footnote{ Wong et al., ``Searching for Life, Mindful of Lyfe’s Possibilities," 7.} Hence, while terrestrial life is based on liquid water, solar energy, and an abundance of six chemical elements (carbon, hydrogen, nitrogen, oxygen, phosphorus, and sulfur – CHNOPS), the definition of lyfe abstracts from these particular circumstances and postulates three general features: the presence of informational driving force (water can be replaced with e.g. ammonia or liquid methane), the energetic driving force (solar can be replaced with e.g. geothermal), and diversity of basic combinatorial elements (CHNOPS can be substituted for different elements, e.g. silicon instead of carbon).\footnote{ Manasvi Lingam and Avraham Leyb, Life in the Cosmos: From Biosignatures to Technosignatures (Cambridge, Massachusetts London, England: Harvard University Press, 2021), 14–22.}

Following this expanded definition of life by Wong, Bartlett, Chen, and Tierney, the concept of habitability also requires an analogous expansion. The authors thus talk about a more generic property of planetary environments they call \textit{genesity}, which includes also ``life-as-we-do-not-know-it". The difference between habitability and genesity boils down for them to the difference between the conditions of \textit{survival} of life forms on the planet on the one hand and the conditions of the planet’s creative ability to initiate the evolution of novel lineages of lyfe-forms (irrespective of their substrate) on the other hand. Interestingly, authors are willing to categorize digital and technical objects as lyfe-forms, hence making them relevant from the standpoint of assessing a planet’s genesity.\footnote{ Wong et al., ``Searching for Life, Mindful of Lyfe’s Possibilities," 10–11.} Likewise, Walker considers advanced technologies such as AI to be forms of life since they stand in the lineage of the information propagation that begins with the origin of terrestrial life and currently finds itself at a major inflection point characterized by the human-aided emergence of the technosphere.\footnote{ Walker, Life as No One Knows It, 233–37.} In turn, one may follow here suggestion by Wong et al. that genesity comes in degrees, and that the planets that support propagation of more diverse and original forms of lyfe have higher genesity than ``just" narrowly habitable worlds.\footnote{ Wong et al., ``Searching for Life, Mindful of Lyfe’s Possibilities," 7–9.} 

Seen from the vantage point of this major conceptual rotation, the relation between sustainability, habitability, and genesity may be stated as follows:

\vspace{1\baselineskip}
\begin{equation}
Sustainability<Habitability<Genesity
\end{equation}

\vspace{1\baselineskip}
\textit{Sustainability} encompasses only those parameters in the topological diagram of planetary history that are conducive to the sustainable development of the human species on Earth. \textit{Habitability} abstracts from the anthropocentric optics of sustainability by encompassing general conditions for the survival of Earth-like life on \textit{any} planet. \textit{Genesity} denotes the most generic category, where the emphasis is placed on parameters conducive to the propagation of lineages of conceivable lyfe-forms. Coupled with the speculative-philosophical implication of the previous chapters – namely that any successful technosphere bends back into the biosphere – maintenance and potential enhancement of genesity then present a vector of planetary history compatible with the Sustainability Solution to Fermi Paradox. The Sustainability Solution at the same time delivers an important corrective that specifies viable trajectories of maintenance and enhancement of any planet’s genesity: The generation of lyfe-forms (including technical objects) should not cross the boundaries delimited by the planet’s biogeochemical metabolisms, which would otherwise decrease the planet’s genesity; in a hypothetical scenario of technosphere suffocating the biosphere, the net value of genesity would be diminished since a mature, planetary-scale assemblage of lyfe-forms would be replaced by its trivialized version. It means that the role of the technosphere is not to replace, but to \textit{expand} the biosphere by the production of novel lyfe-forms. This corrective echoes the idea of any sufficiently advanced technology being indistinguishable from nature – with ``nature" substituted with\textit{ }the planet’s biosphere that contains constructors of the given technosphere.

\section{Conclusion}

In the late 1920s, the first robot ever built in Japan was created in Osaka, designed and manufactured by biologist Makoto Nishimura. The robot’s name was \textit{Gakutensoku }, which means \textit{learning from the laws of nature}. In stark contrast to the Western history of robotics, where the idea of artificial beings has been from the beginning oriented around the imaginaries of hard, manual labor, the first Japanese robot was to embody high moral standards derived from the rules inscribed in nature itself.\footnote{ Yulia Frumer, ``The Short, Strange Life of the First Friendly Robot - IEEE Spectrum," 2020, https://spectrum.ieee.org/the-short-strange-life-of-the-first-friendly-robot.} Given the arguments of this paper, it seems that there is indeed something crucial hidden in the idea of learning from laws of nature – this idea navigates our technological imaginaries towards alignment with the planetary environment and a continuation of more-than-human metabolisms by both biological and technological means, rather than human exceptionalism. In this light, the Sustainability Solution to the Fermi Paradox contains a philosophical takeaway: it tells a story of the convergence of the technosphere with the planet’s pre-existing conditions, rather than the story of replacement or dominance. One can consider this perspective as a radicalization of bio-centric environmental ethics: If one radicalizes bio-centric environmental goals deep enough, one may end up with a protection agenda that involves the technosphere and its generative abilities, to the extent to which it expands a planet’s genesity. Since the planets assume the central role in this normative framework, this paper proposes to follow innovative moral philosophies, such as \textit{planetocentric ethics}.\footnote{ Woodruff T. Sullivan III, ``Planetocentric Ethics: Principles for Exploring a Solar System That May Contain Extraterrestrial Microbial Life," in Encountering Life in the Universe: Ethical Foundations and Social Implications of Astrobiology, ed. William R. Stoeger, Anna H. Spitz, and Chris Impey (Tucson: University of Arizona Press, 2013), 167–77.}

\ \ \ \ This paper is part of a larger research agenda that advocates for embracing astronomical insights in theoretical research across environmental humanities.\footnote{ See also Lukáš Likavčan, ``Another Earth: An Astronomical Concept of the Planet for the Environmental Humanities," Distinktion: Journal of Social Theory 25, no. 1 (January 2, 2024): 17–36, https://doi.org/10.1080/1600910X.2024.2326448.} By surveying the implications of the Sustainability Solution to the Fermi paradox, it engaged with SETI and its philosophical import, arguing that it can be viewed as a speculative exercise with direct consequences for human inhabitation of the Earth. In this vein, research advances in SETI also impact philosophical concepts relevant to environmental humanities, as demonstrated by the reconceptualizations of the technosphere, planetary history, and sustainability through the prism of the Sustainability Solution: 

\vspace{1\baselineskip}
\begin{enumerate}
	\item The concept of technosphere has been critically reconsidered with respect to the dynamics of convergence between technosphere and biosphere (``any sufficiently advanced technology may be indistinguishable from nature"), hypothesized based on the Sustainability Solution’s conjecture about the relation between the lack of observable technosignatures and the unsustainability of exponential growth as the development pattern of planetary communities. The major result of this reconceptualization is the problematization of the analytical import of technosphere as a category denoting some new geological layer – it seems to be more of a transitory armature of the biosphere’s evolution and less of an emerging permanent layer. From this follows a major evolutionary-historical implication: \textit{successful technospheres fold back into the biosphere}.

	\item The concept of planetary history has been expanded into a topological paradigm by abstracting from the case of Earth, focusing instead on the whole space of possible historical pathways available to planets inhabited by sophonts. Limits of biosphere-technosphere coupling present here an important constraint on the parameter space of this topology, resulting in a diagrammatization of planetary history akin to a stellar H-R diagram or diagram of planetary evolutionary trajectories described by Wong and Bartlett.\footnote{ Wong and Bartlett, ``Asymptotic Burnout and Homeostatic Awakening."} This topological perspective welcomes a generic view of the history of the human planetary community on Earth (including its possible future trajectories), which abstracts from the particularities of this planet and this historical moment, in order to plot the overall maneuvering space available for viable inhabitation of planets by sophonts.

	\item The concept of sustainability undergoes two expansions, which propose higher-order normative vectors for the discourses in environmental humanities and practices of climate change mitigation. The first one is habitability, which emphasizes the survival of complex biospheres based on Earth-like organic chemistry, thus problematizing the anthropocentric framework of sustainability. The second one is genesity, which further enlarges the scope of habitability towards non-carbon-based lifeforms or technical objects. Genesity comes in degrees – the higher the planet’s ability to produce novel lyfe-forms, the higher its genesity, which leads to the conclusion that the technosphere’s goal is not to extinguish or overhaul older planetary layers, but to constructively elaborate on top of the planet’s affordances.

\end{enumerate}
\vspace{1\baselineskip}
In terms of further research avenues, the speculative results of this paper first and foremost demand anchoring in theoretical models, such as those developed by Adam Frank.\footnote{ A. Frank et al., ``The Anthropocene Generalized: Evolution of Exo-Civilizations and Their Planetary Feedback," Astrobiology 18, no. 5 (May 2018): 503–18, https://doi.org/10.1089/ast.2017.1671.} This is the case especially with the proposal to think about successful technospheres as those that fold back to biospheres. However, the diagramming of the topological space of planetary histories also needs to first prove its usefulness in actual implementation, whether in exoplanet research, astrobiology, or SETI. Finally, the notion of genesity is largely contingent on one’s willingness to accept the possibility of alternative forms of life, beyond the scope of the terrestrial biosphere. Although different alternative biologies have been proposed and hypothesized, the framework of ``life-as-we-do-not-know-it" is still at best in its pre-paradigmatic stage.\footnote{ Walker, Life as No One Knows It, 29–30.} Nevertheless, the constructive, creative motif intrinsic to the notion of genesity prompts philosophical speculations that can drive normative theories privileging the productive agency of ``endless forms most beautiful and most wonderful."\footnote{ Charles Darwin, On the Origin of Species, ed. Gillian Beer, Rev. ed, Oxford World’s Classics (New York: Oxford University Press, 2008), 360.}

\section*{Acknowledgements}
Special thanks go to Brad Weslake, who commented on earlier versions of this paper during research colloquia at NYU Shanghai. This paper was conceptualized during the author’s visit to ELSI – Earth-Life Science Institute, Tokyo University of Technology, and finalized during a fellowship at the Panel on Planetary Thinking, University of Giessen.

\section*{Author bio}
Lukáš Likavčan is a philosopher focused on emerging technologies, ecology, and astronomy. He currently works as a researcher for the Antikythera program incubated by the Berggruen Institute, as a research affiliate at the Center for AI and Culture, NYU Shanghai, and as a fellow at the University of Giessen's Panel for Planetary Thinking. More info at likavcan.com.

\end{document}